\documentclass[
    ,final            % use final for the camera ready runs
  ]
  {aipproc}

\layoutstyle{6x9}

\begin{document}

\title{Measurement of polarization observables in $\omega$-photoproduction}

% wie HS
% <Replace this text with PACS numbers; choose from this list:
% \texttt{http://www.aip..org/pacs/index.html}>}

\author{H. Eberhardt (for the CBELSA/TAPS collaboration)}{
  address={Rheinische Friedrich-Wilhelms-Universität, Physikalisches Institut,
  D-53115 Bonn, Germany}
}

%\author{<author2>}{
%  address={<common address for author2 and author3>}
%}

%\author{<author3>}{
%  address={<common address for author2 and author3>}
%  ,altaddress={<author1 address>} % additional visiting address
%}

\begin{abstract}
  Near threshold $\omega$ meson photoproduction is studied with the
  CBELSA/TAPS experiment at the ELSA accelerator of Bonn University. Single
  ($\Sigma$, $\Sigma_{\pi}$) as well as double ($G$, $G_{\pi}$, $E$)
  po\-la\-ri\-za\-tion observables are measured using either linear or
  circular polarized photon beams and a longitudinal polarized proton target.
\end{abstract}

\maketitle

%%%%%%%%%%%%%%%%%%%%%%%%%%%%%%%%%%%%%%%%%%%%
%% MAINMATTER
%%%%%%%%%%%%%%%%%%%%%%%%%%%%%%%%%%%%%%%%%%%%

\section{Introduction}

\begin{figure}[htb]
  \includegraphics[width=0.58\textwidth]{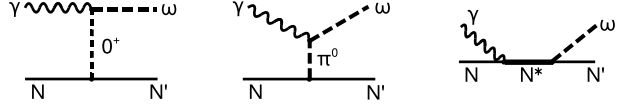}
  \caption{$\omega$ production via t-channel $0^{+}$ (Pomeron) exchange
    (left), t-channel $\pi^{0}$ exchange (middle) and s-channel intermediate
    excitation (right).}
  \label{BildProdMech}
\end{figure}

Baryon spectroscopy provides important information in the non-perturbative
regime of QCD at low and intermediate energies. At these energies constituent
quark models, like the Bonn model \cite{LOR}, try to describe the baryon
spectrum. This and also other quark models have two main conspicuities:
\begin{enumerate}
\item
  The lowest lying resonances $P_{11}(1440)$ (Roper) and $S_{11}(1535)$ are
  not well reproduced. In particular the parity ordering of these states is
  wrong.
\item
  In the higher mass region much more states are predicted than experimentally
  observed.
\end{enumerate}
%\begin{figure}[htb]
%  \includegraphics[width=0.6\textwidth]{Bilder/Nucleon_BonnMod.eps}
%  \caption{The spectrum of Nucleon states obtained in the Bonn model (right
%    side of each column) compared to the measured one (left side of each
%    column). \cite{LOR}}
%  \label{BildBonnMod}
%\end{figure}
The reason for these inconsistencies can be a model deficit. The second may
also be to ex\-pe\-ri\-men\-tal deficits. Most of the experimentally
determined resonances had been extracted from pion-nucleon scattering. Some
resonances may however be hard to observe in these kind of experiments, due to
weak $\pi N$-coupling \cite{CapRob}. One 
possibility to investigate this issue is the photoproduction of $\omega$
mesons off the proton. The production threshold of the $\omega$ meson is
located in the third resonance region, where many of the ``missing'' states
are expected. Due to the fact that the $\omega$ is isoscalar (I=0), the
s-channel production of this meson is only associated with the decay of
$N^{*}$ (I=1/2) states and not the decay of $\Delta^{*}$ (I=3/2) states, which
greatly simplifies the contributing excitation spectrum. However the vector
meson character of the $\omega$ implies that at least 23 observables have to
be measured to disantangle all contributing resonances, instead of 8 in the
pseudoscalar case. It can be hoped however, that fewer than 23 observables
already provide significant constraints.\\
In any case, the measurement of polarization observables will provide
important information about the ``production mechanism'' of the $\omega$ meson
\cite{SAR}. $\omega$ mesons can be prduced via ``diffractive'' scattering in
the t-channel, i.e. via an exchange of a Pomeron (Fig.~\ref{BildProdMech}
(left)) or a Pion (Fig.~\ref{BildProdMech} (middle)). Former analyses, like
the measurement of differential cross sections \cite{BAR}, photon asymmetries
\cite{AJA,KLE} and decay asymmetries (pion asymmetry) \cite{KLE} additionally
show evidence for s-channel contributions (Fig.~\ref{BildProdMech}
(right)). The measurement of further (double) polarization observables will
further clarify the role of s-channel resonances, and it is one goal of
current experiments to investigate this.

% Bild BonnMod weglassen?

\section{Polarization observables}

The cross section for vector meson photo production can be written very
similar to the pseudoscalar case. For the situation of the presented
experiment (using either linear or circular polarized photons and a
longitudinal polarized target) it has the form:
\begin{displaymath}
  \frac{d\sigma}{d\Omega} = \frac{d\sigma_{0}}{d\Omega} ( 1 -
  P_{\gamma,l}\Sigma_{(\pi)}\cos{2\varphi_{(\pi)}} +
  P^{T}_{z}P_{\gamma,l}G_{(\pi)}\sin{2\varphi_{(\pi)}} -
  P^{T}_{z}P_{\gamma,\odot}E )
  \label{Xsec}
\end{displaymath}
where the polarization independent cross section is denoted by $\sigma_{0}$,
the degree of linear polarization by $P_{\gamma,l}$, the degree of circular
polarization by $P_{\gamma,\odot}$ and the degree of longitudinal target
polarization by $P^{T}_{z}$. $\varphi$ is the azimuthal angle of the vector
meson. For the neutral decay of the $\omega$
($\omega\rightarrow\pi^{0}\gamma$), asymmetries for the decay pion can also be
measured. These asymmetries correspond to the observables $\Sigma_{\pi}$ and
$G_{\pi}$ and are obtained by replacing the azimuthal angle of the $\omega$
meson ($\varphi$) by the azimuthal angle of the decay-$\pi^{0}$
($\varphi_{\pi}$). The observables are defined as in Ref. \cite{SAR}.

% polobs ansch ?

\section{Experiment and results}
\begin{figure}[htb]
  \includegraphics[width=0.7\textwidth]{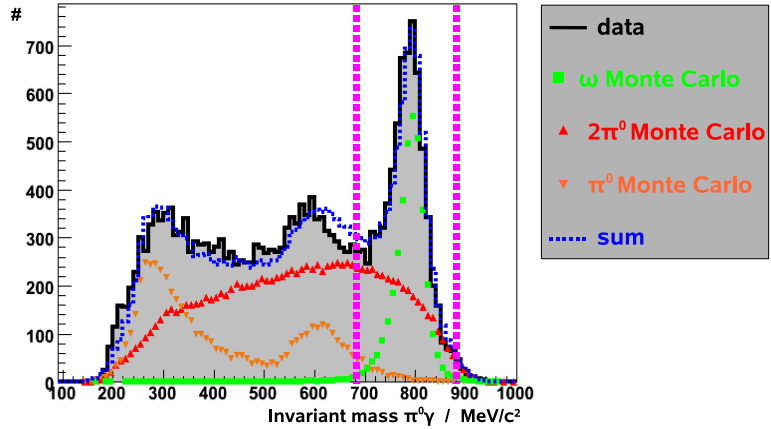}
  \caption{$\pi^{0}\gamma$ invariant mass distribution ($E_{\gamma} = 1300
    -1400~MeV$) compared to the simulated decomposition into $\omega$ signal
    and $\pi^{0}$ and $2\pi^{0}$ background. The cut range for asymmetry
    determination is indicated by the vertical lines.}
  \label{BildInvMass}
\end{figure}

\begin{figure}[htb]
  \includegraphics[width=\textwidth]{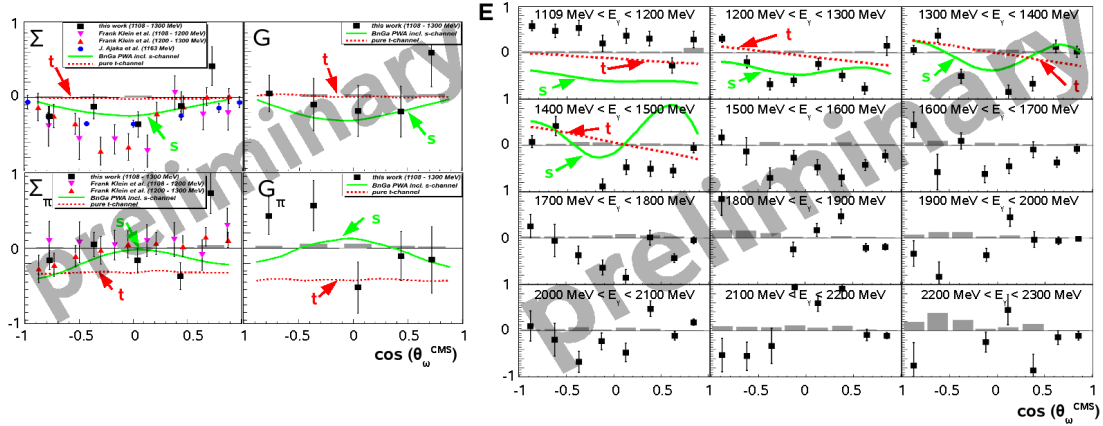}
  \caption{Preliminary results for the polarization observables $\Sigma$,
    $\Sigma_{\pi}$, $G$, $G_{\pi}$ and $E$. The data sample includes only a
    fraction of the complete data set. The observables not measured with a
    polarized target are compared to \cite{AJA} (dots, only $\Sigma$) and two
    energy bins of \cite{KLE} (triangles, $\Sigma$ and $\Sigma_{\pi}$). The
    curves are predictions of the Bonn-Gatchina PWA group \cite{SAR}. The
    solution for only considering t-channel production of the $\omega$ is
    shown by the dotted lines and the solution for dominant contributions from
    s-channel resonances is shown by the continuous lines.}
  \label{BildErgebn}
\end{figure}

The experiment was performed at the tagged photon beam of the ELSA accelerator
in Bonn, using electron beams of $E_{0}~=~2.4$~GeV or
$E_{0}~=~3.2$~GeV. Linear polarized photons, with a maximum degree of
polarization of about 53\%, were produced via coherent
bremsstrahlung from a 500~$\mu$m thick diamond crystal. Circular polarized
photons were generated by bremsstrahlung of longitudinal polarized electrons,
having a degree of polarization of more than 60\% . For double polarization
measurements the Bonn Frozen Spin Target \cite{DUT}, having an average degree
of polarization of about 70\% with relaxation times of about 500 hours, was
used. In order to get rid of systematic errors, two different linear
polarization planes with an azimuthal angular offset of $90^{0}$ were adjusted
and also target and circular polarization directions where flipped at regular
intervals. Surrounding the target was a three layer scintillating fiber
detector, covering the angular acceptance of the Crystal-Barrel
calorimeter. The Crystal-Barrel detector consists of 1230 CsI(Tl) crystals of
16 radiation lengths. It covers the polar angular range of $30^{0}$ -
$150^{0}$. In the forward direction the setup is supplemented by two further
calorimeters, namely the Forward detector and the MiniTAPS detector, where
both of them are equipped with plastic scintillator pads for charge
identification. The whole detector setup covers almost 96\% of the solid angle
around the target. It is well suited for photon detection, i.e. the detection
of the neutral decay of the $\omega$ meson
($\omega\rightarrow\pi^{0}\gamma$).\\
For the event reconstruction three uncharged and one charged reconstructed
particles were demanded. After applying basic kinematic cuts, a $\pi^{0}\gamma$
invariant mass distribution as shown in Fig.~\ref{BildInvMass} was
obtained. Monte Carlo simulation of signal, as well as of background events
showed that the main background channel in the $\omega$ invariant mass range
originated from $2\pi^{0}$ production. These background events can also carry
asymmetries, making it necessary to correct for them.
\\
The polarized target provides a further complipication to the analysis. The
Bonn frozen spin target consists of butanol ($C_{4}H_{10}O$) molecules, where
only the quasi-free hydrogen atoms can be polarized and only the polarization
of these atoms can be measured via NMR techniques. This is the reason why one
has to correct the measured target po\-la\-ri\-za\-tion by the so called
``dilution factor''. The effective dilution factor strongly depends on the
widths of the applied kinematic cuts and on the polar angle of the meson. This
is determined by separate measurements on carbon and hydrogen targets, via
sca\-ling measured carbon and hydrogen spectra to fit the measured butanol
spectra in each energy and polar angular bin\footnote{For example the, out of
  the 4-momentum of the measured meson, calculated proton mass
  spectrum.}. Several methods to obtain the dilution factor are under
investigation.\\
With these corrections one gets preliminary results for the measured
ob\-ser\-va\-bles as shown in Fig.~\ref{BildErgebn}.
\\
The data set for data measured with linear polarized photons is yet incomplete
and thus exhibits large statistical errors. Nevertheless, evidences for
s-channel contributions to the production of the $\omega$ meson, as reported
in \cite{BAR,AJA,KLE}, seem to be further confirmed by the measurement of the
``new'' double polarization observable $E$. Polar angle dependencies of $E$,
particulary at energies close to threshold, have been observed in the data,
whereas \cite{SAR} expects linear dependence on $\cos{\theta_{\omega}^{CMS}}$
if only t-channel contributions were involved.

\section{Summary}
Former analyses \cite{BAR,AJA,KLE} claimed evidence for s-channel contribution
in photo production of $\omega$ mesons in addition to t-channel
production. New (double) polarization data were taken with the CBELSA/TAPS
experiment during the last years (and data taking is ongoing). Preliminary
results on the Beam-Target helicity observable $E$ provides further indication
for s-channel contributions, in accordance to the mentioned analyses.

% summary?

%%%%%%%%%%%%%%%%%%%%%%%%%%%%%%%%%%%%%%%%%%%%%%%%
%% BACKMATTER
%%%%%%%%%%%%%%%%%%%%%%%%%%%%%%%%%%%%%%%%%%%%%%%%

\begin{theacknowledgments}
We acknowledge financial support from the Deutsche Forschungsgemeinschaft
(DFG) within SFB/TR16.
\end{theacknowledgments}



\bibliographystyle{aipproc}   % if natbib is available

%\bibliography{proHE}

%  \typeout{******************************************}
%  \typeout{** Please run "bibtex \jobname" to optain}
%  \typeout{** the bibliography and then re-run LaTeX}
%  \typeout{** twice to fix the references!}
%  \typeout{******************************************}

%\end{document}

%%%%%%%%%%%%%%%%%%%%%%%%%%%%%%%%%%%%%%%%%%%
%% The following lines show an example how to produce a bibliography
%% without the help of the BibTeX program. This could be used instead
%% of the above.
%%%%%%%%%%%%%%%%%%%%%%%%%%%%%%%%%%%%%%%%%%%

\end{document}